\documentclass[twocolumn]{article}
\usepackage{amsmath,amsfonts,amssymb,epsfig,color}

\newcommand{\Spn}[1]{\ensuremath{\mathrm{Sp}( #1 )}}

\newcommand{\so}[1]{\ensuremath{\mathfrak{so}( #1 )}}

\newcommand{\spn}[1]{\ensuremath{\mathfrak{sp}( #1 )}}

\newcommand{\IAS}{isobaric analog $0^+$ state}
\newcommand{\IASs}{isobaric analog $0^+$ states}
\newcommand{\fpg}{\ensuremath{1f_{5/2}2p_{1/2}2p_{3/2}1g_{9/2}} }

\newcommand{\flevel}{\ensuremath{1f_{\frac{7}{2}}} }
\newcommand{\dlevel}{\ensuremath{1d_{\frac{3}{2}}} }

\begin{document}

\title{Microscopic Description of Isospin Mixing Pairing Correlations in
the Framework of an Algebraic Sp(4) Model}
\author{K. D. Sviratcheva$^1$, A. I. Georgieva$^{1,2}$, and J. P. 
Draayer$^1$ \\
$^1$Department of Physics and Astronomy, Louisiana State University,\\
Baton Rouge, Louisiana 70803, USA \\
$^2$Institute of Nuclear Research and Nuclear Energy,\\
Bulgarian Academy of Sciences, Sofia 1784, Bulgaria}
\date{\today}

\maketitle

\begin{abstract}
We explore isospin mixing beyond that due to  the Coulomb interaction 
in the framework of
an exactly solvable microscopic \spn{4} algebraic approach. 
Specifically, we focus on the
isospin  non-conserving part  of the {\it pure} nuclear pairing 
interaction. The  outcome of this
study shows the significance of the pairing charge dependence and its 
role in mixing
isospin multiplets of pairing-governed \IASs~in light and  medium 
mass nuclei, especially in nuclei
with equal numbers of protons and neutrons. The model reveals 
possible,  but still extremely weak,
non-analog $\beta $-decay transitions and estimates their relative 
strengths within a shell closure. 
\end{abstract}

\section{Introduction}

A major simplification of the pairing problem is achieved if
one assumes a charge independent nuclear interaction (the proton-proton
$pp$ interaction and the neutron-neutron $nn$ interaction are equal to the
isospin $T =1$  proton-neutron $pn$ interaction).  The latter
comprises a quite reasonable approximation and
consequently most of isovector (isospin 1) pairing studies have been
carried under the assumption of isospin invariance.  However,
``the problem of broken symmetry is one of general significance
in  nuclear (e.g.
\cite{Ekman04}) and elementary particle (e.g.
\cite{BarnesCL03}) physics" \cite{BohrMottelson} (Vol. I, p.37), 
which has been of long-standing interest
\cite{Wigner57,GoswamiChen,Auerbach72,BertschW73,BohrMottelson,
TownerHH77,Lawson,BenensonK79} and may be associated with  novel
and interesting physics
\cite{DobaczewskiH95,OrmandBrown,SagawaGS96,CivitareseReboiroVogel,
NavratilBO97,HardyT02,Lisetskiy,BesC04,AbergHMR04,
MichelNP04,PetroviciSRF05,AlvarezRGS05}.
Experimental results clearly reveal the existence of  isospin mixing
\cite{Hagberg,Savard95}. An increase in isospin mixing towards medium
mass nuclei has been
detected in novel high-precision experiments
\cite{Cottle99,Garrett01,Piechaczek03,Farnea03,Ekman04}, which
continue to push the
exploration of unstable nuclei  with the advent of advanced
radioactive beam facilities.

  The isospin symmetry in nuclei is slightly violated by  the
electromagnetic interaction, mainly the Coulomb repulsion between nucleons
\cite{BohrMottelson,Auerbach72,DobaczewskiH95}. Another source of 
mixing probability is
the isospin  non-conserving part of the nuclear Hamiltonian, which 
includes effects due
to the proton-neutron  mass difference ($\Delta m/m=1.4 \times 
10^{-3}$) and small charge dependent
components in the  strong nucleonic interaction that appear to 
be associated with
the electromagnetic  structure of the nucleons \cite{Auerbach72}. An 
analysis of the $^1S$ scattering
in the $pn$  system and the low-energy
$pp$ scattering  lead to the estimate that the pure nuclear 
interaction  between protons and
neutrons ($V_{pn}^{T =1}$) in $T =1$ states are more attractive than 
the force between
the protons ($V_{pp}$) by $2\%$, $|V_{pn}^{T =1}-V_{pp}|/V_{pp} \sim 2\%$
\cite{Henley66}.
Furthermore, after the Coulomb energy is taken into account the 
discrepancy in the
isobaric-multiplet energies is bigger for the seniority zero  levels 
as compared to
higher-seniority  states indicating the presence of a short range 
charge dependent
interaction \cite{Lawson}. Indeed, the $J=0$ pairing correlations 
have been recently
shown to have an overwhelming dominance in the isotensor energy 
difference within
isobaric multiplets \cite{ZukerLMP02}, which manifests  itself in the 
charge dependent
$T=2$ nature of the pairing interaction.

 The aforementioned findings set the need for a charge dependent
microscopic description of $J=0$ pairing correlations.
For this reason we employ a simple but powerful
group-theoretical model \cite{SGD03,SGD04IAS}, which is based on the 
\spn{4} algebra
(isomorphic to \so{5} \cite{Hecht,Ginocchio,EngelLV96}).
A comparison with experimental data demonstrates that the
\Spn{4} model provides a reasonable description of the  pairing-governed
\IASs\footnote{The lowest among these states include ground states 
for even-even nuclei and only some
($N\approx Z$) odd-odd nuclei, as well as, for example, low-lying 
$0^+$ states in odd-odd nuclei that
have the same isospin as the ground state of a semi-magic even-even 
isobaric neighbor with fully-paired
protons (or neutrons). } in light and  medium mass nuclei, where 
protons and neutrons occupy the same
shell \cite{SGD03,SGD03stg,SGD04IAS}.  The \Spn{4} model is precisely 
suitable for the
microscopic modeling of the pairing interaction and its isospin 
violation in \IASs~ because it
naturally extends the isospin invariant nuclear interaction to 
incorporate isospin non-conserving
forces, while it retains the \Spn{4} dynamical  symmetry of the 
Hamiltonian.  Hence it provides a
straightforward scheme for estimating the significance of the isospin 
mixing due to pairing
correlations without the need for carrying out large-dimensional 
matrix diagonalizations. Strong
isospin breaking in pair formation, if found, implies a significant 
presence of isospin admixture among
the seniority-zero
\IASs~including $0^+$ ground states. This in turn will affect the 
predictive power of precise studies
of superallowed  $0^+ \rightarrow 0^+$ Fermi $\beta
$-decay transitions.  Laboratory and theoretical investigations of 
such transitions provide reliable
tests of isospin mixing (see \cite{TownerH02} for a review). In 
addition, the results of examinations
on isospin mixing  are essential to another challenging problem; 
namely, when compared to the decay
rate for  purely leptonic muon decay, the estimate for the nuclear 
Fermi $\beta $ decay rate furnishes
a  precise test of the unitary condition of the 
Cabbibo-Kobayashi-Maskawa matrix \cite{CKM63and73}
under the assumption of the three-generation standard particle model 
(for a review of this subject, see
\cite{TownerH03}).

\section{Isospin Mixing of the Isobaric Analog $0^+$ States}

The \Spn{4} model reflects the symplectic dynamical symmetry of
\IASs~ \cite{SGD04IAS} determined by the strong nuclear interaction.
The weaker Coulomb
interaction breaks this symmetry and significantly complicates the
nuclear  pairing problem.
This is why, in our investigation we adopt a sophisticated
phenomenological Coulomb  correction  to the
experimental energies such that a nuclear system can be regarded as
if there is no  Coulomb
interaction between its constituents. The {\it Coulomb corrected}
experimental energy,
$E_{\exp }$, for given valence protons $N_{+1}$ and neutrons $N_{-1}$
is adjusted to be
\begin{eqnarray}
E_{\exp }(N_{+1},N_{-1})&=&E^C_{\exp }(N_{+1},N_{-1})- E^C_{\exp }(0,0)
\nonumber
\\ &+&V_{Coul}(N_{+1},N_{-1}),
\label{EexpCoul}
\end{eqnarray}
where\footnote{To avoid confusion we mention that in
(\ref{EexpCoul}) the energies are
assumed positive for bound states; $V_{Coul}$ is also defined positive.}
$E^C_{\exp }$ is the total measured energy including the Coulomb
energy \cite{AudiWapstra,Firestone}, $E^C_{\exp }(0,0)$ is
the binding energy of the core, and $V_{Coul}(N_{+1},N_{-1})$ is the 
Coulomb correction for a nucleus
with mass $A$ and $Z$
protons taken relative to the core $V_{Coul}(N_{+1},N_{-1})=
V_{Coul}(A,Z)-V_{Coul}(A_{core},Z_{core})$. The recursion formula for
the $V_{Coul}(A,Z)$
Coulomb energy is derived in \cite {RetamosaCaurier} with the use of
the Pape and Antony formula \cite{PapeAntony88}
\begin{equation}
\fontsize{8}{10pt}\selectfont
V_{Coul}(A,Z)=\left\{
\begin{array}{r}
V_{Coul}(A,Z-1)+1.44\frac{(Z-1/2)}{A^{1/3}}-1.02\\
Z>Z_{s} \\
V_{Coul}(A,Z+1)-1.44\frac{(Z+1/2)}{A^{1/3}}+1.02\\
Z<Z_{s},
\end{array}
\right.
\label{VCoul}
\end{equation}
where $Z_{s}=A/2$ for $A$ even or $Z_{s}=(A+1)/2$ for $A$ odd. When $Z=Z_{s}$
the Coulomb potential is given by
\begin{equation}
\fontsize{8}{10pt}\selectfont
V_{Coul}(A,Z_{s})=\left\{
\begin{array}{c}
0.162Z_{s}^{2}+0.95Z_{s}-18.25\qquad Z_{s}\leq 20 \\
0.125Z_{s}^{2}+2.35Z_{s}-31.53\qquad Z_{s}>20.
\end{array}
\right.
\end{equation}
The Coulomb corrected energies (\ref{EexpCoul}) should reflect
solely the nuclear properties
of the many-nucleon systems.

Assuming charge independence of the nuclear force, the general
isoscalar Hamiltonian with Sp(4) dynamical symmetry, which consists
of one- and two-body terms  and conserves the number of particles, can be
expressed  through the Sp(4) group generators,
\begin{eqnarray}
H_{0} =&-G\sum _{i=-1}^{1}\hat{A}^{\dagger }_{i}
\hat{A}_{i}-\frac{E}{2\Omega} (\hat{T}
^2-\frac{3\hat{N}}{4 })
\nonumber \\
&-C\frac{\hat{N}(\hat{N}-1)}{2}-\epsilon  \hat{N},
\label{clH0}
\end{eqnarray}
where $\hat{T}^2=\Omega \{ \hat{T}_+,\hat{T}_-\}+\hat{T}_0^2$ and
$2\Omega $ is the shell dimension for a given nucleon type. The
generators $\hat{T}_{\pm}$
and $\hat{T}_{0}$ are the valence isospin operators,
$\hat{A}^{(\dagger ) }_{0,+1,-1}$ create (annihilate) respectively a
proton-neutron ($pn$)
pair, a  proton-proton ($pp$) pair or a neutron-neutron
$(nn)$ pair of total angular momentum $J^{\pi}=0^+$ and isospin
$T=1$, and $\hat{N} =
\hat{N}_{+1}+\hat{N}_{-1}$ is the total number of valence particles
with an eigenvalue $n$.
The $G,E$ and $C$ are interaction strength parameters  and
$\epsilon >0$ is the Fermi level energy (see Table I in
\cite{SGD04IAS} for estimates). The isospin
conserving Hamiltonian
(\ref{clH0}) includes an isovector ($T=1$) pairing interaction
($G\geq  0 $ for attraction)
and a diagonal isoscalar ($T=0$) force, which is related to a
symmetry term ($E$).
The two-body  model
interaction includes proton-neutron and like-particle pairing plus symmetry
terms and contains a non-negligible  implicit portion of the
quadrupole-quadrupole interaction
\cite{cmpRealInt05}. Moreover, the
\Spn{4} model interaction itself, which relates to the whole energy
spectrum rather than to a single $J^\pi =0^+$ $T=1$ state, was found to be
quite strongly correlated ($0.85$) with the realistic CD-Bonn+3terms
interaction \cite{PopescuSVN05} in the $T=1$ channel and with an overall
correlation of $0.76$ with the realistic GXPF1 interaction 
\cite{HonmaOBM04} for
the \flevel orbit \cite{cmpRealInt05}. In short, the relatively simple
\Spn{4}  model seems to be a reasonable approximation that reproduces 
especially
that part of  the interaction that is responsible for shaping pairing-governed
\IASs.

Charge dependent but charge
symmetric\footnote{The charge asymmetry between the
$pp$  and $nn$ interactions  is found to be small; namely, less than $1\%
$ \cite{Baumgartner66}.}
nucleon-nucleon interaction
($V_{pp}=V_{nn} \ne V_{pn}$)  brings into the nuclear
Hamiltonian a  small isotensor component (with zero third isospin
projection so that the
Hamiltonian commutes with
$T _0$). This is achieved in the framework of the \Spn{4} model by
introducing the two
additional terms,
\begin{equation}
H_{\text{IM}}=-F \hat{A}^{\dagger }_{0}\hat{A}_{0}, \qquad
H_{\text{split}}=-D(\hat{T} _{0}^2-\frac{\hat{N}}{4}),
\label{HINC}
\end{equation}
to the isospin invariant model Hamiltonian (\ref{clH0}) in a way that
the Hamiltonian
\begin{equation}
H=H_{0}+H_{\text{IM}}+H_{\text{split}}
\label{clH}
\end{equation}
possesses \Spn{4} dynamical symmetry. The interaction strength
parameters $F$ and $D$ (\ref{HINC}) determined in an optimum fit over a
significant number of nuclei (total of 149)
\cite{SGD03} yield non-zero values (see Table I in
\cite{SGD04IAS} for estimates). As expected from observations,
for the \dlevel level the interaction strengths of all $pn$, $pp$ and
$nn$ pairing are found to be almost equal ($T$ is a good quantum
number),
$F/\Omega = 0.007$, and they differ for the \flevel and for the
\fpg  shells, with the $pn$ isovector strength being more attractive,
$F>0$.
The full Hamiltonian (\ref{clH}) yields
quantitative  results that are superior than the ones with $F=0$ and $D=0$; for
example, in  the case of the
\flevel level the variance between the model and experimental energies of the
lowest
\IASs~ increases by $85\%$ when the $D$ and $F$ interactions are turned off.
For the  present investigation the parameters in (\ref{clH0}) along 
with $F$ and
$D$ (\ref{HINC}) are not varied as their values were fixed to be physically
valid and to yield reasonable energy
\cite{SGD03,SGD04IAS} and fine structure \cite{SGD03stg} reproduction for light
and medium mass nuclei with valence protons and  neutrons  occupying the same
shell. For these nuclei in the mass range $32\le A\le 100$,  the
pairing-governed \IASs~are well described, but still approximately, by the
eigenvectors of the effective Hamiltonian (\ref{clH}) in a basis of 
fully-paired ($pp$, $pn$ and $nn$ $T=1$ pairs) $0^+$ states (Table \ref{tab:classifCa}).
\begin{table*}[th]
\center{
\caption {
Classification scheme of even-$A$ nuclei in the \flevel  shell. The
shape of the table is symmetric with respect to the sign of $T_0$ and
$n-2\Omega $. The operators  shown in brackets (and their Hermitian
conjugates) generate
transitions in the action space spanned by the $|n,T,T_0 \rangle$ 
isospin eigenstates that are
linear combinations of the fully-paired basis states.}
\begin{tabular}{c|cccccc}
\hline \hline
& \multicolumn{6}{c}{Isospin projection $T_0$} \\
$n$ & \multicolumn{1}{c}{2} & 1& 0 & -1 &
\multicolumn{1}{c}{-2} &
\multicolumn{1}{c}{-3} \\
\hline
0 & &  &
\begin{tabular}{c}
$_{20}^{40}$Ca$_{20}$ \\
$\left| 0,0,0 \right)$
\end{tabular}
&  &  &
\\
2 & &
\begin{tabular}{c}
$_{22}^{42}$Ti$_{20}$ \\
$\left| 2,1,1 \right)$
\end{tabular}
&
\begin{tabular}{c}
$_{21}^{42}$Sc$_{21}$  \\
$\left| 2,1,0 \right)$
\end{tabular}
&
\begin{tabular}{c}
$_{20}^{42}$Ca$_{22}$   \\
$\left| 2,1,-1 \right)$
\end{tabular}
&
\multicolumn{1}{c}{} &  \\
4 & $\swarrow $ & \multicolumn{1}{c}{
\begin{tabular}{c}
$_{23}^{44}$V$_{21}$  \\
$\left| 4,2,1 \right)$ \\
\hspace{0.1cm}
\end{tabular}
} &
\begin{tabular}{c}
$_{22}^{44}$Ti$_{22}$ \\
$\left| 4,2,0 \right)$ \\
$\left| 4,0,0 \right)$
\end{tabular}
&
\begin{tabular}{c}
$_{21}^{44}$Sc$_{23}$  \\
$\left| 4,2,-1 \right)$ \\
\hspace{0.1cm}
\end{tabular}
&
\multicolumn{1}{c}{
\begin{tabular}{c}
$_{20}^{44}$Ca$_{24}$  \\
$\left| 4,2,-2 \right)$ \\
\hspace{0.1cm}
\end{tabular}
}
& \multicolumn{1}{c}{}   \\
6 & $\cdots $ &
\multicolumn{1}{c}{
\begin{tabular}{c}
$_{24}^{46}$Cr$_{22}$ \\
$\left| 6,3,1 \right)$ \\
$\left| 6,1,1 \right)$
\end{tabular}
} &
\begin{tabular}{c}
$_{23}^{46}$V$_{23}$ \\
$\left| 6,3,0 \right)$ \\
$\left| 6,1,0 \right)$
\end{tabular}
&
\begin{tabular}{c}
$_{22}^{46}$Ti$_{24}$ \\
$\left| 6,3,-1 \right)$ \\
$\left| 6,1,-1 \right)$
\end{tabular}
&
\multicolumn{1}{c}{
\begin{tabular}{c}
$_{21}^{46}$Sc$_{25}$ \\
$\left| 6,3,-2 \right)$ \\
\hspace{0.1cm}
\end{tabular}
} &
\multicolumn{1}{c}{$\searrow $ {\tiny $(A^{\dagger }_{-1})$}}  \\
8 & $\leftarrow ^{ {\tiny (T_+)}}$  &
\begin{tabular}{c}
$_{25}^{48}$Mn$_{23}$   \\
$\left| 8,4,1 \right)$ \\
$\left| 8,2,1 \right)$ \\
\hspace{0.1cm}
\end{tabular}
&
\begin{tabular}{c}
$_{24}^{48}$Cr$_{24}$   \\
$\left| 8,4,0 \right)$ \\
$\left| 8,2,0 \right)$ \\
$\left| 8,0,0 \right)$
\end{tabular}
&
\begin{tabular}{c}
$_{23}^{48}$V$_{25}$   \\
$\left| 8,4,-1 \right)$ \\
$\left| 8,2,-1 \right)$ \\
\hspace{0.1cm}
\end{tabular}
&
\begin{tabular}{c}
$_{22}^{48}$Ti$_{26}$  \\
$\left| 8,4,-2 \right)$ \\
$\left| 8,2,-2 \right)$ \\
\hspace{0.1cm}
\end{tabular}
&
\multicolumn{1}{c}{$\dots $} \\
10 & $\vdots $
&$\vdots $ &
$\downarrow$ {\tiny ($A^{\dagger }_0$)} &\multicolumn{1}{c}{$\vdots
$} & $\vdots $ & \multicolumn{1}{c}{$\swarrow $ {\tiny $(A^{\dagger
}_{+1})$}}  \\
\hline \hline
\end{tabular}
\label{tab:classifCa}
}
\end{table*}

While the second interaction ($H_{\text{split}}$) in (\ref{HINC})
takes into account only the
splitting of the isobaric analog energies, the first correction
induces small isospin mixing
(IM). The isospin mixing interaction (\ref{HINC}) does not account
for the entire interaction
that mixes states of same angular momentum and parity but  different
isospin values. It only describes a possible $\Delta T=2$ mixing between
\IASs~due to a pure nuclear pairing interaction. While the extent
of such isospin admixing is expected to be smaller than the total mixing due to
isospin non-conserving terms
\cite{TownerHH77,OrmandBrown,SagawaGS96,NavratilBO97,TownerH02}, it
may influence precise model calculations depending on the importance of the
charge dependence in pairing correlations.

The question regarding how strong individual isospin non-conserving
nuclear interactions are
[such as (\ref{HINC})] still remains open -- there are no sharp
answers at the present level
of experimental results and microscopic theoretical interpretations.
It  is only their overall
contribution that is revealed by the free nucleon-nucleon data
\cite{Henley66} to be slightly (by $2\%$) more attractive in the $pn$
$T =1$ system than  the
$pp$ one. Within the framework of the \Spn{4} model, the charge
dependence of the pure
nuclear interaction can be estimated through the comparison of the
$T_0=0$ two-body model
interaction [(\ref{clH}) with $\varepsilon =0$] relative to the
$T_0=1$ in the $T  =1$
multiplets, which, for example in the \flevel level, is on average
$\sim 2.5\%$. In addition, the \Spn{4} model reproduces reasonably well the
$c$-coefficient in the well-known isobaric multiplet mass equation
\cite{Wigner57,BenensonK79,Ormand97}
\begin{equation} a+bT_0+cT_0^2,
\label{IMME}
\end{equation}
for the binding energies of isobaric analogs (of the  same mass number
$A$, isospin $T$, angular momentum $J$, etc.), where the coefficient
$c$ ($b$) depends on the isotensor (isovector) component of the nuclear
interaction [i.e., of  rank 2 (1) with  respect to isospin `rotations'].
The requirement that the  coefficients of (\ref{IMME}) are well reproduced is 
essential for the isospin
non-conserving models
\cite{TownerHH77,OrmandBrown,ZukerLMP02}, which has been achieved in
\cite{TownerHH77}  by increasing (approximately by $2 \%$) of all the
$T=1$ $pn$ matrix
elements relative to the $nn$ ones and which has lead to a conclusion
in \cite{ZukerLMP02}
that the  isotensor nature of the nuclear interaction is dominated by
a $J=0$ pairing term.
In agreement with experiment, the $c$-coefficients in the
\Spn{4} model were found to be negative and very close to zero for $T=1$
multiplets in the \flevel shell. Their average  relative to the binding
energy of the valence nucleons differs from the corresponding
experimental value by only 0.3\%. These estimations do not aim to
confirm the  charge-dependence, which is very difficult at this level of
accuracy compared to the  broad energy range considered in the model for
nuclei with masses $32 \le A \le 100$. Nonetheless,  it reflects the
fingerprints of the experimental data in the properties of the model
interaction (\ref{clH}). 

\subsection{Non-Analog $\beta $-Decay Transitions}

For a superallowed Fermi $\beta$-decay transition ($0^+ \rightarrow
0^+$) the $ft$ comparative
lifetime is nucleus-independent according to the
conserved-vector-current (CVC) hypothesis and
given by
\begin{equation}
ft=\frac{K}{G_V^2|M_F|^2 }, \quad K =2\pi ^3\hbar \ln 2
\frac{(\hbar c)^6}{(m_e c^2)^5},
\end{equation}
where $K/(\hbar c)^6=8.120 270 (12)\times 10^{-7}$ GeV$^{-4}$s ($m_e$
is the mass of the
electron) and $G_V$ is the vector coupling  constant for nuclear
$\beta $ decay (see for
example \cite{OrmandBrown}). $M_F$ is  the Fermi matrix element $\left<
\text{F}|\sqrt{2\Omega} T _{\pm } |\text{ I} \right>$  between a
final (F) state with isospin
projection $T _0^{\text{F}}$ and an initial (I)  states with $T
_0^{\text{I}}$ in a decay
generated by the raising (for
$\beta ^-$ decay) and lowering ($\beta ^+$) isospin transition
operator\footnote{The factor
of $2\Omega $ appears  due to the normalization of the basis
operators adopted in the \spn{4}
algebraic model.} $\sqrt{2\Omega}T _{\pm }$, which in the framework
of our model is given as
\begin{equation}
|M_F|^2=2\Omega
|\left< \text{F};n(\tilde T)T_0\pm 1|T _{\pm }
|\text{I};n(\tilde T)T_0 \right>|^2,
\end{equation}
where $\left| n(\tilde{T})T_0 \right\rangle$ are the 
eigenvectors of the total
Hamiltonian (\ref{clH}) with an almost good isospin $\tilde{T}$ 
quantum number.
\begin{figure}[th]
\centerline{\epsfxsize=3.0in\epsfbox{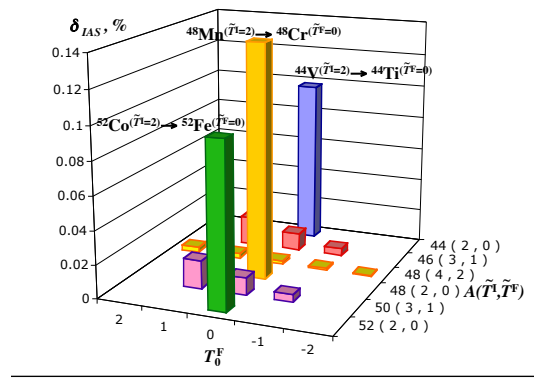}}
\caption{
\Spn{4} model estimate for the $\delta _{IAS}$ isospin mixing 
correction [\%]
(\ref{deltaIAS}) to Fermi $\beta $-decay transition matrix elements between
\IASs~ of almost good isospin $\tilde{T}$ for the nuclei in the 
\flevel level ($F/\Omega =0.072$).}
\label{CaBetaDecays}
\end{figure}
Typically, the isospin impurity caused by isospin non-conserving
forces in nuclei is estimated
as a correction to the Fermi matrix element
$|M_F|^2$ of the superallowed $\tilde T$ analog $0^+
\rightarrow 0^+$ transition,  $\delta _C= 1- |M_F|^2/\left\{\tilde T
(\tilde T +1) -T
_0^{\text{F}}T _0^{\text{I}} \right\}$. For more than two-state
mixing, the degree of isospin
admixture between \IASs~ should be estimated using  the normalized
transition matrix element
between non-analog (NA) states (e.g. \cite{TownerH02}),
\begin{equation}
\delta_{IAS}=\frac{|M_F^{\text{NA}}|^2}{\left\{ \tilde T (\tilde T
+1) -T _0^{\text{F}}T
_0^{\text{I}}\right\}},
\label{deltaIAS}
\end{equation}
where $\tilde T$ is the almost good isospin of the parent nucleus (Figure
\ref{CaBetaDecays} for the \flevel level). The small mixing of 
the $0^+$ isospin eigenstates
from different isospin multiplets reflects very small but nonzero 
$|M_F^{NA}|^2$  matrix elements for
non-analog $\beta ^{\pm} $ decay transitions (Figure
\ref{CaBetaDecays} and first column of Table \ref{tab:betaCa}).
In  short, the theoretical \Spn{4}  model suggests the possible existence, albeit  highly
hindered, of $\Delta T =2$ non-analog $\beta $-decay transitions.
\begin{table}[th]
\caption{Non-analog $\beta $-decay transitions to energetically
accessible $0^+$ states under consideration for nuclei in  the
\flevel level along with the parameter-free ratio of the first-order 
isospin mixing
$\delta ^{(1)}_{IAS}$ relative to $\delta ^{(1)}_{IAS}$ of the 
$_{23}^{44}\text{V}^{(2)} \rightarrow
_{22}^{44}\text{Ti}^{(0)}$ decay (denoted as $\delta ^{(1)*}_{IAS}$) 
in the framework of the
\Spn{4} model. (There are no available experimental values for comparison.)}
\center{
\begin{tabular}{llll}
\hline \hline
\multicolumn{3}{c}{$\beta$-decay} & $\frac{\delta 
^{(1)}_{IAS}}{\delta ^{(1)*}_{IAS}}$ \\ [1ex]
$^A_ZX^{(\tilde T _X)}$ & $\rightarrow $ & $ ^{\hspace{0.35cm}
A}_{Z-1}Y^{(\tilde T _Y)}$ &   \\ [1ex]
\hline
$_{23}^{44}\text{V}^{(2)} $ & $\rightarrow $ & $ 
_{22}^{44}\text{Ti}^{(0)}$ & 1.000 \\ [1ex]
\hline
$_{25}^{46}\text{Mn}^{(3)} $ & $\rightarrow $ & $ _{24}^{46}\text{Cr}^{(1)}$
     &0.173 \\ [1ex]
$_{24}^{46}\text{Cr}^{(3)} $ & $\rightarrow $ & $ _{23}^{46}\text{V}^{(1)}$
     &0.115 \\ [1ex]
$_{23}^{46}\text{V}^{(3)} $ & $\rightarrow $ & $ _{22}^{46}\text{Ti}^{(1)}$
     &0.043 \\ [1ex]
\hline
$_{27}^{48}\text{Co}^{(4)} $ & $\rightarrow $ & $ _{26}^{48}\text{Fe}^{(2)}$
     &0.034 \\ [1ex]
$_{26}^{48}\text{Fe}^{(4)} $ & $\rightarrow $ & $ _{25}^{48}\text{Mn}^{(2)}$
     &0.030 \\ [1ex]
$_{25}^{48}\text{Mn}^{(4)} $ & $\rightarrow $ & $ 
_{24}^{48}\text{Cr}^{(2)}$ &0.020 \\[1ex]
$_{24}^{48}\text{Cr}^{(4)} $ & $\rightarrow $ & $ 
_{23}^{48}\text{V}^{(2)}$ &0.011 \\[1ex]
$_{23}^{48}\text{V}^{(4)} $ & $\rightarrow $ & $ 
_{22}^{48}\text{Ti}^{(2)}$ &0.004 \\[1ex]
$_{25}^{48}\text{Mn}^{(2)} $ & $\rightarrow $ & $ 
_{24}^{48}\text{Cr}^{(0)}$ &1.452 \\[1ex]
\hline
$_{27}^{50}\text{Co}^{(3)} $ & $\rightarrow $ & $ _{26}^{50}\text{Fe}^{(1)}$
     & 0.173 \\ [1ex]
$_{26}^{50}\text{Fe}^{(3)} $ & $\rightarrow $ & $ _{25}^{50}\text{Mn}^{(1)}$
     &0.115 \\ [1ex]
$_{25}^{50}\text{Mn}^{(3)} $ & $\rightarrow $ & $ _{24}^{50}\text{Cr}^{(1)}$
     &0.043 \\ [1ex]
\hline
$_{27}^{52}\text{Co}^{(2)} $ & $\rightarrow $ & $ _{26}^{52}\text{Fe}^{(0)}$
     &1.000 \\ [1ex]
\hline \hline
\end{tabular} }
\label{tab:betaCa}
\end{table}

\begin{figure}[th]
\centerline{\epsfxsize=3.0in\epsfbox{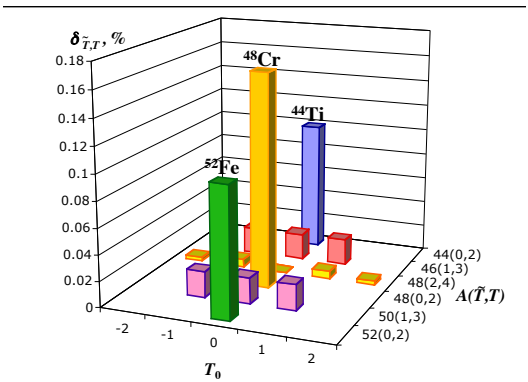}}
\caption{
\Spn{4} model estimate for the mixing overlap [\%] of $0^+$ states under consideration of almost
good isospin $\tilde{T}$ with the states of definite isospin for the 
nuclei in the \flevel level
($F/\Omega =0.072$).}
\label{mixingOvlpf7}
\end{figure}

In general, the
$\delta _{IAS}$ correction may be very different than the order of 
the $\delta _{\tilde{T },T }$
overlap quantity
\begin{equation}
\delta _{\tilde{T },T }=\left| \left\langle n,T ,T_0| n(\tilde{T 
})T_0 \right\rangle \right|
^{2}*100[\%]
\label{delta}
\end{equation}
of the $\left| n(\tilde{T})T_0 \right\rangle$ nuclear states 
with the isospin eigenvectors
(Figure \ref{mixingOvlpf7}). This is because in decays the degrees 
of isospin mixing between
non-analog states within both the parent and daughter nuclei are 
significant.  As it is expected, the
$\delta  _{\tilde{T },T } $ isospin mixing increases as
$Z$ and $N$ approach one another and towards the  middle of the 
shell. Although the isospin admixture
is negligible for light nuclei in the $ j=3/2$ orbit ($\delta 
_{\tilde T=0,T=2}=0.0001$ for $^{36}$Ar, $F/\Omega =0.007$), it is clearly bigger for the $ j=7/2$ 
level (Figure \ref{mixingOvlpf7}),
yet less than $0.17\%$.

The analysis of the results for the \flevel orbit shows that the 
mixing between \IASs~
(which is at least
$\Delta T =2$ mixing) is on average $0.006 \%$ excluding even-even
$N=Z$ nuclei (Figure \ref{CaBetaDecays}). This is on the order of a 
magnitude less than the
mixing of the first excited
$0^+$ non-analog  state due to isospin non-conserving interaction,
which is  typically about
$0.04 \%$ for the
\flevel level \cite{Hagberg,TownerH02}. In addition, it is smaller
than possible Gamow-Teller
transitions, $<0.02\%$ for the nuclei in the
\flevel shell \cite{Hagberg}, that are found substantially larger
with increasing mass number
$A$ \cite{Hamamoto93,HardyT02,Piechaczek03}. This makes
$\delta _{IAS}$ mixing very difficult to be detected especially when
the isospin-symmetry
breaking correction ($\delta _C$) to analog Fermi  matrix elements in
this level is on the
order of a percent \cite{DobaczewskiH95,HardyT02}.

Despite the lack of experimental data, a rough estimate for
the order of the
$\delta _{IAS} $ mixing induced by $H_{IM}$ (\ref{HINC}) in the
\Spn{4} model can be obtained
in  comparison to other types of mixing corrections that are
measured or calculated. For example, the $\beta $-decay
transition from  the ground state of
$^{46}$V to the first excited
$0^+$ non-analog state in $^{46}$Ti at $E_{0^+_1}=2.61$ MeV yields an
experimental correction
of $0.053\%$ \cite{Hagberg}, which is also reproduced by theoretical
calculations \cite{TownerH02}.
The first excited
$0^+$ $\tilde T=2$ state in $^{46}$Ti that is an isobaric analog to
the ground state of
$^{46}$Ca lies at $E_{\tilde T =2}=13.36$ MeV (as predicted by the
\Spn{4} model
\cite{SGD04IAS}) and its mixing into the $\tilde T=0$ ground state
should yield an isospin
symmetry-breaking correction on the order of:
\begin{equation}
\delta _{IAS}\sim \left(\frac{E_{0^+_1}}{E_{\tilde T
=2}}\right)^2 0.053 \% \sim 0.0020\%,
\label{estimIAS}
\end{equation}
which is about the values the \Spn{4} model yields for $\Delta T=2$
non-analog transition
between $^{46}$V and $^{46}$Ti (Figure \ref{CaBetaDecays}). Clearly,
such a comparison is approximate with respect to the high
accuracy of the isospin
mixing effects. Yet the results show consistency
with other theoretical
calculations and are found not to contradict reasonable limits set up
by experimental evidences.

Not surprising, the largest values for the $\delta _{IAS} $
correction are observed for $\Delta T=2$
$\beta ^{\pm}$-decays to energetically accessible $0^+$ ground states of
even-even
$N=Z$ nuclei (Figure \ref{CaBetaDecays}). While for these decays $\delta
_{IAS}$ is extremely
small, namely less than $0.14 \%$, as expected for the  contribution
of the higher-lying $0^+$
states \cite{TownerH02}, it is comparable to the  order of
isospin-symmetry breaking
corrections for the \flevel orbit that are typically taken  into account
\cite{TownerH02}. The reason is that for the even-even $N=Z$
nuclei the second-lying \IASs~ are situated relatively low due to a significant
$pn$  interaction (Figure \ref{CaBetaDecays}). As an  example, 
one finds that for the decay
to  the $^{48}$Cr ground
state  $\delta _{IAS}$ may be only about 5 times  smaller [proportional to the
ratio in the energies squared as in (\ref{estimIAS})] than an average
$\Delta T=1$
isospin-symmetry breaking correction in the \flevel orbit and  takes
the latter to be around
$0.6\%$ \cite{TownerH02}. Indeed, the \Spn{4} model yields $\delta
_{IAS}=0.14328\%$ for $_{25}^{48}\text{Mn}^{(2)} \rightarrow 
_{24}^{48}\text{Cr}^{(0)}$.

Above all, the $\delta _{IAS} $ results in Figure \ref{CaBetaDecays} 
clearly show the
overall pattern and the order of significance of the isospin mixing under
consideration. This is evident within the first-order approximation in terms of
the $F$ parameter ($F \ll 1$) of $\delta _{IAS} $ (Table 
\ref{tab:betaCa}), which for
\flevel deviates on average by only 2\% from its exact  calculations 
in Figure \ref{CaBetaDecays}. The
$\delta ^{(1)}_{IAS} $ isospin mixing correction is then proportional 
to $F^2$ and one finds out that
its order of  magnitude remains the same for large variations of the 
$F$ parameter of more than 60\%.
In addition, greater $F$ values are not very likely because the 
$\delta _{IAS} $ estimates
(Figure \ref{CaBetaDecays}) fall close below an upper limit, which 
does not contradict experimental and
theoretical results for other types of isospin mixing. It is worth
mentioning that while the energies of the lowest \IASs~ determined directly the
parameters of the model interaction, a quite good reproduction of the
experimental higher-lying
\IAS~energies followed without any parameter adjustment \cite{SGD04IAS}. This
outcome is important because the energy difference between two \IASs~ within a
nucleus directly affects the degree of their mixing.

Moreover, in this first-order approximation the
ratio of any two isospin corrections within a shell, where the 
strengths of the
effective interaction are assumed fixed, is independent of the 
parameters of the model interaction.
This implies that such a ratio does not reflect at all the 
uncertainties of the interaction strength
parameters but rather it is  characteristic of the relative strength 
of both decays. We choose to
compare $\delta ^{(1)}_{IAS} $  for different $\Delta T=2$ 
$\beta$-decays to the isospin mixing
correction, denoted by $\delta ^{(1)*}_{IAS} $, of the decay between 
nuclear isobars with $n=4$ valence
particles [such as the $_{23}^{44}\text{V}^{(2)} \rightarrow 
_{22}^{44}\text{Ti}^{(0)}$ decay for the
\flevel orbit (Table
\ref{tab:betaCa})]
due to the relative simplicity of these nuclear systems. The $\delta 
^{(1)}_{IAS}/\delta ^{(1)*}_{IAS}$
ratio then identifies the  decay, for which the maximum isospin 
mixing correction is expected in the
\flevel orbit, namely
$_{25}^{48}\text{Mn}^{(2)} \rightarrow  _{24}^{48}\text{Cr}^{(0)}$, and as well
as the amount by which $\delta _{IAS} $ of the other possible non-analog decays
is relatively suppressed (Table
\ref{tab:betaCa}). For example, the $\delta _{IAS} $ correction for the
$_{23}^{44}\text{V}^{(2)}
\rightarrow  _{22}^{44}\text{Ti}^{(0)}$ decay is around 1.5 times smaller than
the maximum one and it is around 8 times smaller for the
$_{25}^{46}\text{Mn}^{(3)} \rightarrow  _{24}^{46}\text{Cr}^{(1)}$ 
decay. Such a ratio quantity
exhibits a general trend of increasing
$\delta _{IAS} $ isospin mixing with $Z$ within same isospin multiplets and as
well it reveals enhanced $\Delta T=2$ decays to the ground state of   even-even
$N=Z$ nuclei with increasing $\delta _{IAS} $ towards the middle of the shell.

Furthermore, the ratio retains its behavior for the non-analog
$\beta $ decays between nuclei with the same valence proton and neutron numbers
as in Table \ref{tab:betaCa} but occupying the
\fpg major shell (Table \ref{tab:betaNi}). Therefore, among the 
non-analog $\beta $ decays for the
$A=60-64$ isobars with valence protons and neutrons in the \fpg shell the
$\delta _{IAS} $ isospin mixing of the
$_{33}^{64}\text{As}^{(2)} \rightarrow  _{32}^{64}\text{Ge}^{(0)}$ decay is
expected to be the largest with a tendency of a further increase towards the
middle of the shell. While the decay mentioned above
exhibits isospin mixing twice stronger than the one for the
$_{31}^{60}\text{Ga}^{(2)}
\rightarrow  _{30}^{60}\text{Zn}^{(0)}$ decay, the other $A=60-64$ 
$\beta $ decays are up to 110 times
slower (Table \ref{tab:betaNi}). In addition, one needs to 
calculate only the
isospin mixing for the simplest case of four valence nucleons, for 
then the order of significance
of $\delta _{IAS} $ for the other
$A=62-64$ decays follow directly from the estimations presented in 
Table \ref{tab:betaNi}.
\begin{table}[th]
\caption{Non-analog $\beta $-decay transitions to energetically
accessible $0^+$ states under consideration for $A=60-64$ nuclei in  the
the upper $fp$ shell along with the parameter-free ratio of the 
first-order isospin mixing
$\delta ^{(1)}_{IAS}$ relative to $\delta ^{(1)}_{IAS}$ of the
$_{31}^{60}\text{Ga}^{(2)} \rightarrow _{30}^{60}\text{Zn}^{(0)}$ 
decay (denoted as $\delta
^{(1*)}_{IAS}$) in the framework of the \Spn{4} model. (There are no 
available experimental values for
comparison.)}
\center{
\begin{tabular}{llll}
\hline \hline
\multicolumn{3}{c}{$\beta$-decay} & $\frac{\delta ^{(1)}_{IAS}}{\delta 
^{(1)*}_{IAS}}$ \\ [1ex]
$^A_ZX^{(\tilde T _X)}$ & $\rightarrow $ & $ ^{\hspace{0.35cm}
A}_{Z-1}Y^{(\tilde T _Y)}$ &   \\ [1ex]
\hline
$_{31}^{60}\text{Ga}^{(2)} $ & $\rightarrow $ & $ 
_{30}^{60}\text{Zn}^{(0)}$ & 1.000 \\ [1ex]
\hline
$_{33}^{62}\text{As}^{(3)} $ & $\rightarrow $ & $ _{32}^{62}\text{Ge}^{(1)}$
     &0.233 \\ [1ex]
$_{32}^{62}\text{Ge}^{(3)} $ & $\rightarrow $ & $ _{31}^{62}\text{Ga}^{(1)}$
     &0.156 \\ [1ex]
$_{31}^{62}\text{Ga}^{(3)} $ & $\rightarrow $ & $ _{30}^{62}\text{Zn}^{(1)}$
     &0.058 \\ [1ex]
\hline
$_{35}^{64}\text{Br}^{(4)} $ & $\rightarrow $ & $ _{34}^{64}\text{Se}^{(2)}$
     &0.081 \\ [1ex]
$_{34}^{64}\text{Se}^{(4)} $ & $\rightarrow $ & $ _{33}^{64}\text{As}^{(2)}$
     &0.072 \\ [1ex]
$_{33}^{64}\text{As}^{(4)} $ & $\rightarrow $ & $ 
_{32}^{64}\text{Ge}^{(2)}$ &0.049 \\[1ex]
$_{32}^{64}\text{Ge}^{(4)} $ & $\rightarrow $ & $ 
_{31}^{64}\text{Ga}^{(2)}$ &0.026 \\[1ex]
$_{31}^{64}\text{Ga}^{(4)} $ & $\rightarrow $ & $ 
_{30}^{64}\text{Zn}^{(2)}$ &0.009 \\[1ex]
$_{33}^{64}\text{As}^{(2)} $ & $\rightarrow $ & $ 
_{32}^{64}\text{Ge}^{(0)}$ &2.301 \\[1ex]
\hline
& & \vdots & \\ [1ex]
\hline \hline
\end{tabular} }
\label{tab:betaNi}
\end{table}

Even though the strength of the isospin mixing interaction may 
differ between
different model spaces, the ratio, $\delta 
^{(1)}_{IAS}/\delta ^{(1)*}_{IAS}$, turns out to be of the same
order for both the \flevel level (Tables \ref{tab:betaCa}) and the upper
$fp$ shell (Table
\ref{tab:betaNi}). In addition, the ratio strongly correlates for 
both shells retaining the same
behavior as one goes from one model space to the other. In short, the 
significance of the isospin mixing
caused by a charge dependent
$J=0$ pairing correlations is evident from Table \ref{tab:betaCa} for 
the \flevel
level and continues the same trend for the upper $fp$ shell (Table 
\ref{tab:betaNi}).

\begin{figure}[th]
\centerline{\epsfxsize=3.0in\epsfbox{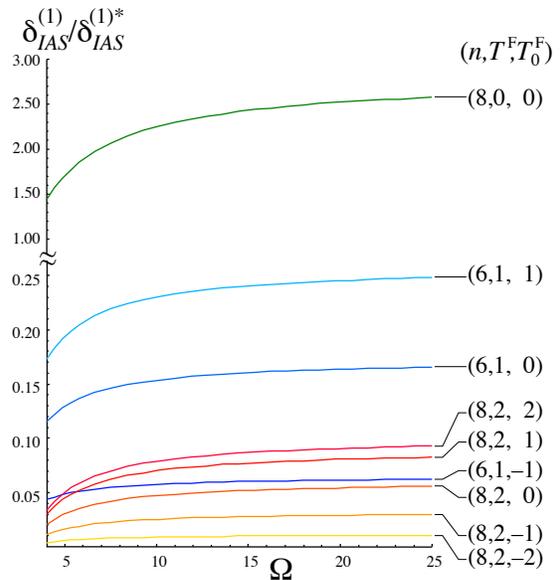}}
\caption{
\Spn{4} model estimate for the first-order mixing correction 
for the $\Delta T=2$
non-analog $\beta $-decays between $0^+$ states under consideration 
for $n$ valence nucleons occupying a
model space of size $\Omega $ (e.g., $\Omega =4$ for \flevel and 
$\Omega =11$ for $\fpg$) relative to
the simplest $n=4$ $\Delta T=2$ decay to the ground states of the 
$N=Z$ nucleus.}
\label{deltaVsOmega}
\end{figure}

Indeed, the relative strength of the first-order isospin mixing 
correction for non-analog Fermi $\beta
$-decays under consideration varies smoothly with the size of the 
model space (Figure
\ref{deltaVsOmega}). All the $\Delta T=2$ decays relative to the one 
with four valence nucleons
need an isospin mixing correction to the transition matrix elements 
that increases with the occupation
space. Such an increase however keeps all of the decays under 
consideration suppressed relatively to
the $n=4$ decay (with the largest $\delta ^{(1)}_{IAS}$ still about 
four times smaller than $\delta
^{(1)*}_{IAS}$). An exception is the $\Delta T=2$ decay to the ground 
state of the $N=Z$
$n=8$ nucleus, which is significantly faster for any $\Omega $ space 
size. This decay becomes $2.5$
times faster in the pair-boson limit of very large $\Omega$, which is 
an increase of $77.6\%$
compared to $\Omega=4$. While the $n=6$ nuclear systems exhibit an 
increase of only $43.8\%$, the
isospin mixing correction increases 1.8 times for all the $n=8$ 
decays to a daughter nucleus of almost
good isospin $\tilde T =2$.
In short, the \Spn{4} model allows one to 
easily estimate the relative
strength of different decays within a shell and hence to identify the 
fastest decay as well as the ones
that can be readily neglected in precise isospin mixing calculations.

\section{Conclusions}

Isospin mixing induced by a short-range
charge dependent nuclear interaction is described microscopically 
within the framework of a
group-theoretical approach based on the \Spn{4} dynamical symmetry. The
\Spn{4} model interaction incorporates the main driving forces, 
including $J=0$ pairing correlations and
implicit quadrupole-quadrupole term, that shape the nuclear pairing-governed
\IASs~in the \flevel  level where the \Spn{4} Hamiltonian correlates strongly
with realistic interactions.

Empirical evidence such as scattering analysis and the 
coefficient related to the isotensor
part  of a general non-conserving force, $c$, reveals the charge 
dependence of the $J=0$ pairing
correlations. Indeed, the slightly stronger proton-neutron pairing 
interaction than the like-particle
(proton-proton or neutron-neutron) pairing interaction came 
out of the \Spn{4} analysis in
a   quite good reproduction of the 
energies of the lowest \IASs~ and the
$c$-coefficient for a wide-range nuclear systems.
The freedom allowed in the algebraic model by introducing additional 
non-conserving forces reflects
the symmetries observed in light nuclei (good  isospin) and the 
comparatively larger
symmetry-breaking as expected in medium-mass  nuclei.
The isospin-symmetry breaking due to coupling  of \IASs~ in nuclei was
estimated to be extremely small for nuclei in the \dlevel and \flevel 
orbitals.  However, the $N=Z$
even-even light and medium mass nuclei are
an exception. For these nuclei, strong pairing correlations,
including a significant $pn$
interaction, are  responsible for the existence of comparatively
larger isospin mixing,
although the latter is still at least an order of a magnitude
smaller than
the overall isospin admixture in the ground state. The results
also show that
a variation of more than 60\% in the $F$ isospin mixing parameter is
required to reduce the present $\delta _{IAS}$ results by an order of a
magnitude.

The analysis also shows that there is a trend of
increasing isospin mixing  between \IASs~due to a charge dependent 
$J=0$ pairing
interaction towards the middle of the shell and for $\Delta T=2$ decays to the
ground state of an even-even $N=Z$ daughter nucleus. Such behavior is 
free of the
uncertainties in the strength parameters of the interaction
and is adequate for larger multi-$j$ shell domains such as
$\fpg$. For nuclei with valence protons and neutrons occupying
the \flevel
level the strongest non-analog decay is identified to be
$_{25}^{48}\text{Mn}^{(2)}
\rightarrow  _{24}^{48}\text{Cr}^{(0)}$ with the $\delta _{IAS}$ isospin mixing
correction being 1.5 to 300 times smaller for the rest of the decays. 
In the upper $fp$ shell
among the decays between nuclei with mass $A=60-64$ the 
$_{33}^{64}\text{As}^{(2)} \rightarrow
_{32}^{64}\text{Ge}^{(0)}$ decays is found to be the fastest while 
the isospin mixing decreases for the
other decays 2 to 250 times.

In general, relative to the simplest decay between isobars with four 
valence nucleons, the
parameter-free isospin mixing corrections for the same number of 
valence nucleons increase with the size of
the occupation space. Hence, such a ratio is larger in the \fpg major 
shell than in the
\flevel single-$j$ orbit. The trend observed allows for 
estimates for the isospin mixing
correction to
$\beta $-decay matrix elements between isobars with four to eight 
valence particles but filling other
major shells. Moreover, the present study provides for the order of 
significance of isospin mixing due
to pairing correlations for the decays under consideration if only 
the $\delta _{IAS}$ mixing
correction for the simplest decay is provided by model calculations 
or experimental observations.
Hence large and negligible isospin mixing corrections are easily 
identified. For example, in the \flevel
orbit the \Spn{4} model yields $\delta _{IAS}=0.098\%$ for 
$_{23}^{44}\text{V}^{(2)} \rightarrow
_{22}^{44}\text{Ti}^{(0)}$ and hence for the fastest $_{25}^{48}\text{Mn}^{(2)}
\rightarrow  _{24}^{48}\text{Cr}^{(0)}$ decay it is $\delta _{IAS}=0.143\%$.

In short, the charge dependence of the
nuclear force, being a very challenging problem, yields results,
based on a simple
group-theoretical approach, that  are qualitatively as well as 
quantitatively consistent with
the observations.
The \spn{4}  algebraic model
yields an estimate for the  decay rates of possible
non-analog $\beta$-decay transitions due to a pure strong
interaction, which, though few of them may
affect slightly precise calculations, are not
expected to comprise the dominant contribution to the  isospin-symmetry
breaking correction tested in studies of superallowed Fermi $\beta$-decay
transitions.

\section*{Acknowledgments}
This work was supported by the US National Science  Foundation, Grant
Number 0140300. The
authors thank Dr. Vesselin G. Gueorguiev for his computational
{\small MATHEMATICA}\ programs
for non-commutative algebras.

\end{document}